\documentclass{ws-ijmpe}

\begin{document}

\markboth{W. Trautmann and H.H. Wolter}{Elliptic Flow and the Symmetry Energy}

\catchline{}{}{}{}{}

\title{ELLIPTIC FLOW AND THE SYMMETRY ENERGY\\ 
AT SUPRA-SATURATION DENSITY
}

\author{\footnotesize W. TRAUTMANN}

\address{GSI Helmholtzzentrum f\"{u}r Schwerionenforschung GmbH\\
Planckstr. 1, D-64291 Darmstadt, Germany\\
w.trautmann@gsi.de}

\author{H.H. WOLTER}

\address{Fakult\"{a}t f\"{u}r Physik, Universit\"{a}t M\"{u}nchen\\
Am Coulombwall 1, D-85748 Garching, Germany\\
hermann.wolter@lmu.de}

\maketitle

\begin{history}
\received{(received date)}
\revised{(revised date)}
\end{history}

\begin{abstract}
The elliptic flow in collisions of neutron-rich heavy-ion systems at intermediate 
energies emerges as an observable sensitive to the strength of the symmetry energy 
at supra-saturation densities. First results obtained by comparing ratios or 
differences of neutron and hydrogen flows with predictions of transport models
favor a moderately soft to linear density dependence, consistent with ab-initio 
nuclear-matter theories. 
Comprehensive data sets of high accuracy can be expected to improve our knowledge of the 
equation of state of asymmetric nuclear matter. 
\end{abstract}

\section{Introduction}

Heavy-ion reactions at sufficiently high energy represent the only means for compressing 
nuclear matter in laboratory experiments and for studying the nuclear equation of state (EoS)
at supra-saturation densities.\cite{dani02} From the extensive search for observables 
suitable for probing the brief high-density phase of the collision, 
collective flows and sub-threshold 
production of strange mesons have appeared as most useful. The present consensus that a soft
EoS, corresponding to a compressibility $K \approx 200$~MeV and including 
momentum dependent interactions, best describes the
high-density behavior of symmetric nuclear matter is based on studies of flow and kaon 
production within the framework of transport theory.\cite{dani02,sturm01,fuchs01}

In recent years, motivated by the impressive progress made in observing properties of neutron 
stars and in understanding details of supernova explosion scenarios, the EoS of neutron-rich
asymmetric matter has received increasing attention.\cite{lattprak07,lipr08,ditoro10} 
The symmetry energy, 
which is the difference between the energy per nucleon of neutron matter and of symmetric 
matter, is considered to be one of the biggest unknowns in this context. We have precise information 
on the symmetry energy near saturation density from the knowledge of nuclear masses.
For densities below saturation, investigations of heavy ion
reactions in the Fermi energy regime have constrained the symmetry energy
considerably,\cite{tsang09} and the
importance of clustering for the symmetry energy at very low density has recently been 
demonstrated.\cite{nato10,horo06,typel10}
At super-saturation density, however, the symmetry energy is still largely unknown for several 
reasons. Phenomenological forces, even though well constrained near saturation, yield 
largely diverging results if they are extrapolated to higher densities.\cite{brown00,fuchs06}
Many-body calculations with realistic potentials face the problem that three-body forces and 
short-range correlations are not sufficiently constrained at higher densities 
at which their importance 
increases.\cite{subedi08,xuli10,steiner12} Even the magnitude of the kinetic contribution,
related to the nuclear Fermi motion and considered as principally understood, is possibly
modified by a redistribution of nucleon momenta due to short-range correlations in 
high-density nuclear matter.\cite{carb11}
\vskip -0.6cm 

\begin{figure}[!htb]
 \leavevmode
 \begin{center}
  \includegraphics[angle=270,width=0.50\columnwidth]{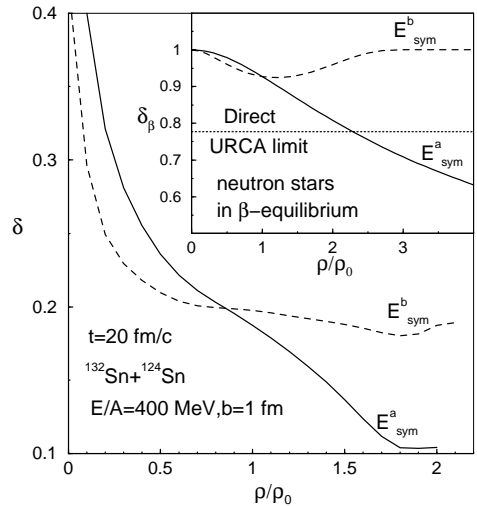}
 \end{center}
\vskip -0.1cm  
  \caption{Isospin asymmetry $\delta = (\rho_n -\rho_p)/\rho$ as a function of the 
normalized density $\rho/\rho_0 $ 
at time $t = 20$~fm/$c$ in the $^{132}$Sn + $^{124}$Sn reaction with 
a stiff ($E^a_{\rm sym}$) and with a soft ($E^b_{\rm sym}$) density dependence of the 
nuclear symmetry energy. 
The corresponding correlation for neutron stars in $\beta$-equilibrium is shown in
the inset (reprinted with permission from Ref.\protect\cite{li02}; 
Copyright (2002) by the American Physical Society).}

\label{fig:li_prl02fig2}
\end{figure}
\vskip -0.3cm 

The appearance of the symmetry energy in many aspects of nuclear structure and reactions
would seem to imply that constraints of various kinds should be available. 
However, quantities as, e.g.,
the thickness of the neutron skin in heavy nuclei or the isospin transport 
in reactions of isospin-asymmetric nuclei in the Fermi-energy domain are predominantly 
sensitive to the strength of the symmetry term at densities below 
the saturation value.\cite{lipr08,tsang09,baran05,klimk07,tamii11}
Extrapolations of parameterizations to higher density raise the question of up to where
they remain realistic. This emphasizes the need for more direct high-density probes.
As in the case of symmetric nuclear matter, collective flows and sub-threshold particle
production are obvious candidates.

A strong motivation for exploring the information contained in isotopic flows was provided 
by Bao-An Li when he pointed to the parallels in the density-dependent isotopic compositions 
of neutron stars and of the transient systems formed in collisions of neutron-rich nuclei 
as a function of the EoS input used in the calculation.\cite{li02} 
Figure~\ref{fig:li_prl02fig2} illustrates the remarkable fact that 
the same physical laws are confidently applied to objects differing by 18 orders of magnitude 
in linear scale or 55 orders of magnitude in mass. It will allow us to infer properties of 
these exotic astrophysical objects from data obtained in laboratory experiments. 
The main difficulty is the comparatively small asymmetry of available nuclei.
Symmetry effects are, therefore, always small relative to the dominating isoscalar forces 
which one hopes will cancel in  
differences or ratios of observables between isotopic partners.
The observable proposed by Li is the so-called differential directed flow which is the
difference of the multiplicity-weighted directed flows of neutrons and protons. Directed
flow describes the rapidity dependence of the mean in-plane transverse momenta of observed 
reaction products. 

A further encouragement was provided by transport model calculations
according to which the elliptic flow of free neutrons and protons responds significantly
to variations of the parameterization of the symmetry energy.\cite{russotto11}
Elliptic flow relates to the azimuthal anisotropy of particle emissions.
This has motivated a reanalysis of the FOPI/LAND data for $^{197}$Au + $^{197}$Au collisions 
at 400 MeV/nucleon, collected many years ago and used to demonstrate the existence of 
neutron squeeze-out in this energy regime.\cite{leif93,lamb94} 
Squeeze-out refers to a dominant out-of-plane
emission of particles, relative to in-plane emission, and is considered as evidence for the
pressure buildup in the collision zone. The analysis favors a moderately soft to linear 
density dependence of the symmetry energy.\cite{russotto11} 

This finding had a particular significance, 
in spite of a large statistical uncertainty. Rather different conclusions, suggesting
either a super-stiff or super-soft behavior of the symmetry energy, had previously
been reached in analyses of the $\pi^-/\pi^+$ production ratios, measured by the
FOPI Collaboration for the same $^{197}$Au + $^{197}$Au reaction,\cite{reis07} 
with different transport models.\cite{ferini05,xiao09,feng10} 
In particular, the super-soft result has initiated 
a broad discussion of how it might be reconciled with other observations as, e.g., 
observed properties of neutron stars.\cite{xiao09,baoan11,wen09} 
The FOPI/LAND elliptic-flow data were found to be inconsistent with this extreme 
assumption also when compared to recent QMD transport-model calculations, 
fairly independent of particular choices made for other model parameters.\cite{cozma11}

The obvious need to improve the statistical accuracy beyond that of the existing
data set has initiated a dedicated measurement of
collective flows in collisions of $^{197}$Au + $^{197}$Au as well as of $^{96}$Zr + $^{96}$Zr 
and $^{96}$Ru + $^{96}$Ru which has been carried out in 2011 with the LAND\cite{LAND} detector 
coupled to a subset of the CHIMERA\cite{CHIMERA} detector array.\cite{s394}
In the following sections, the present situation will be described in more detail.

\begin{figure}[!htb]
 \leavevmode
 \begin{center}
  \includegraphics[width=0.69\columnwidth]{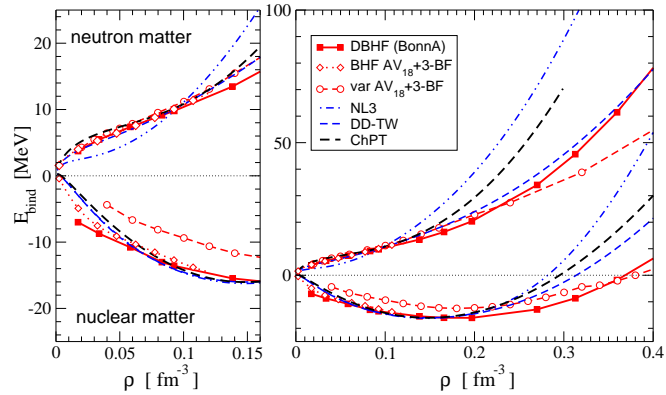}
 \end{center}
\vskip -0.1cm  
  \caption{EoS in nuclear matter and neutron matter.  
BHF/DBHF and variational calculations with realistic forces are compared to  
phenomenological density functionals NL3 and DD-TW and to 
ChPT. The left panel zooms the low density range  
(from Ref.\protect\cite{fuchs06}, 
reprinted with kind permission from Springer Science+Business Media).
}
\label{fig:fuchs06}
\end{figure}

\section{Present Knowledge of the Symmetry Energy}

Our knowledge of the symmetry energy is originally based on nuclear masses whose dependence 
on the isotopic composition is reflected by the symmetry term in the Bethe-Weizs\"{a}cker 
mass formula. A density dependence is already indicated by the use of separate bulk and 
surface terms in refined mass-formulae. In the Fermi-gas model, it is given by a proportionality
to $(\rho / \rho_0)^{\gamma}$ with an exponent $\gamma = 2/3$,
where $\rho_0 \approx 0.16~{\rm nucleons/fm}^{3}$ is the saturation density.
This so-called kinetic 
contribution to the symmetry energy, however, amounts to only about 1/3 of the symmetry term
of approximately 30 MeV for nuclear matter at saturation. The major contribution
is given by the potential term reflecting properties of the nuclear forces.

\begin{figure}[!htb]
 \leavevmode
 \begin{center}
  \includegraphics[width=0.7\columnwidth]{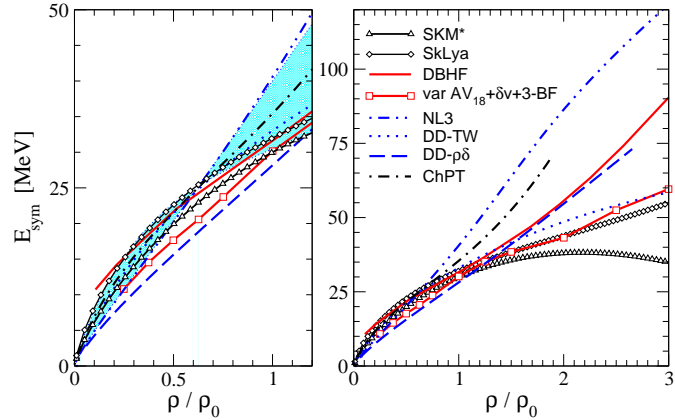}
 \end{center}
\vskip -0.1cm  
  \caption{Symmetry energy as a function of density as predicted by different  
models. The left panel zooms the low density range up to saturation. The full lines
represent the DBHF and variational approaches using realistic forces   
(from Ref.\protect\cite{fuchs06}, 
reprinted with kind permission from Springer Science+Business Media).
}
\label{fig:fuchs06b}
\end{figure}

Nuclear many-body theory has presented us with a variety of predictions for the nuclear
equation of state.\cite{dani02,fuchs06,baldo04} 
The examples shown in Fig.~\ref{fig:fuchs06} for the two cases of 
symmetric nuclear matter and of pure neutron matter demonstrate that, overall, the results
are quite compatible among each other, except for densities exceeding saturation at 
which the predictions diverge. 
The symmetry energy $E_{\rm sym}$ can be defined as the coefficient of the quadratic 
term in an expansion of the energy per particle in the asymmetry  
$\delta = (\rho_n-\rho_p)/\rho$, where $\rho_n, \rho_p,$ and $\rho$
represent the neutron, proton, and total densities, respectively,
\begin{equation}
E/A(\rho,\delta) = E/A(\rho,\delta = 0) + E_{\rm sym}(\rho)\cdot \delta^2 + \mathcal{O}(\delta^4).
\label{eq:e_sym}
\end{equation}

\noindent In the quadratic approximation, the symmetry energy is the difference 
between the energies of symmetric matter ($\delta = 0$) and neutron matter ($\delta = 1$).
Also the symmetry energy diverges at high density, as expected from 
Fig.~\ref{fig:fuchs06}, while most empirical models coincide near or slightly below saturation, 
the density range at which constraints from finite nuclei are valid (Fig.~\ref{fig:fuchs06b}).

The behavior of the symmetry energy at very low densities is not correctly described 
in homogeneous mean-field approaches which do not include clustering effects.
This is a topic of considerable current interest with good agreement being reached
by experimental and theoretical investigations.\cite{nato10,horo06,typel10} The uncertainty at 
supra-saturation density in phenomenological models may be considered as the expected 
consequence of extrapolations leading beyond the range at which the chosen parameterizations 
are effectively tested.\cite{brown00} 

In calculations using realistic forces fitted to two- and three-nucleon data, the 
uncertainty is mainly related to the short-range behavior of the nucleon-nucleon force
and, in particular, to the three-body and tensor forces.\cite{subedi08,xuli10,steiner12} 
The three-body force has been shown to make an essential but quantitatively small contribution 
to the masses of light nuclei.\cite{wiringa02} 
The extrapolation of the partly phenomenological terms used there to higher densities is, 
however, highly uncertain.\cite{xuli10} 
The general effect of including three-body forces in the calculations is a 
stiffening of the symmetry energy with increasing density.\cite{burgio08,hebe10}
Short-range correlations become also increasingly important at higher densities;
results from very recent new experiments will, therefore, have a 
strong impact on predictions for high-density nuclear matter.\cite{subedi08,carb11}
Increasingly precise data from neutron-star observations may further provide
useful constraints.\cite{steiner10}

At higher energies, the momentum dependence of the nuclear forces
becomes important.\cite{lipr08,ditoro10,giordano10,feng12} 
It is well known that nuclear mean fields are momentum dependent, as seen, e.g., 
in the energy dependence of the nuclear optical potential. The dominating effect 
is in the isoscalar sector but there is also an important isovector momentum
dependence. It manifests itself as an energy dependence of the isospin-dependent part
of the optical potential
but can also be expressed in terms of a difference of the effective masses of protons and
neutrons.\cite{fuchs06}. Even the ordering of these effective masses is still an open
problem. It has, moreover, been shown that the 
effective mass differences and the asymmetry dependence of the EoS are both influencing
particle yields and flow observables, and that additional observables will be needed
to resolve the resulting ambiguity.\cite{ditoro10,giordano10,feng12}
\vskip -0.2cm 

\begin{figure}[!htb]
 \leavevmode
 \begin{center}
  \includegraphics[width=0.75\columnwidth]{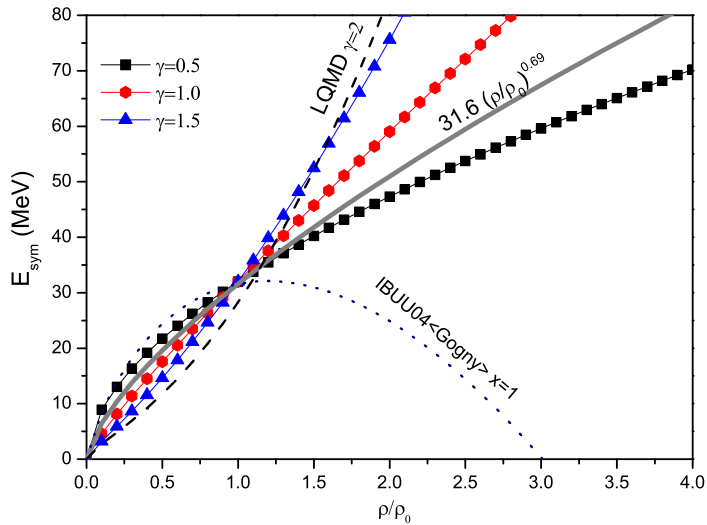}
 \end{center}
\vskip -0.1cm  
  \caption{Parameterizations of the nuclear symmetry energy as used in transport codes: 
three parameterizations of the potential term used in the 
UrQMD (Ref.\protect\cite{qli05}) 
with power law coefficients $\gamma = 0.5, 1.0$, and 1.5 (lines with symbols as indicated), 
the result with $\gamma = 0.69$ obtained from analyzing
isospin diffusion data with the IBUU04 (full line, Ref.\protect\cite{lichen05}),
and the super-soft and stiff parameterizations obtained from analyzing the $\pi^-/\pi^+$ 
production ratios with the IBUU04 (dotted line, Ref.\protect\cite{xiao09}) and the LQMD 
(dashed line, Ref.\protect\cite{feng10}) transport models
(from Ref.\protect\cite{guo12},
reprinted with kind permission from Springer Science+Business Media).
}
\label{fig:params}
\end{figure}

Transport theories needed for calculating the temporal evolution of nuclear reactions
often use simplified parameterized descriptions of the composition-dependent part of 
the nuclear mean field.  
In the UrQMD of the group of Li and Bleicher,\cite{qli05} 
the potential part of the symmetry energy is defined with two parameters, 
the value at saturation density, usually taken as 22 MeV in their calculations, 
and the power-law coefficient $\gamma$ describing the dependence on
density, 
\begin{equation}
E_{\rm sym} = E_{\rm sym}^{\rm pot} + E_{\rm sym}^{\rm kin} 
= 22~{\rm MeV} \cdot (\rho /\rho_0)^{\gamma} + 12~{\rm MeV} \cdot (\rho /\rho_0)^{2/3}.
\label{eq:pot_term}
\end{equation}

\noindent In other codes the nuclear potential of Das {\it et al.} 
with explicit momentum dependence in the isovector sector is used.\cite{das03} 
There, as in the IBUU04 developed by the groups of Li and Chen,\cite{lipr08,lichen05} 
the density dependence of the symmetry energy is characterized by a parameter $x$ 
appearing in the potential expressions.
Examples of these parameterizations and of results obtained from the analysis of experimental
reaction data are given in Fig.~\ref{fig:params}. The stiff ($E^a_{\rm sym}$) and 
soft ($E^b_{\rm sym}$) density dependences of Fig.~\ref{fig:li_prl02fig2} correspond 
approximately to the cases $\gamma = 1$ and $x = 1$ shown here.

Parameterizations of this kind have the consequence that, once the symmetry energy at the 
saturation point is fixed, a single value at a different density or, alternatively, 
the slope or curvature at any density will completely determine the 
parameterization. Measurements of a variety of observables in nuclear structure and reactions 
have, therefore, been used to obtain in this way results for the density dependence of the 
symmetry energy. They are often
expressed in the form of the parameter $L$ which is proportional to the slope 
at saturation, 
\begin{equation}
L = 3\rho_0 \cdot dE_{\rm sym}/d\rho |\rho_0.
\label{eq:l}
\end{equation} 

\noindent Most results with their errors fall into the interval 20~MeV $\le L \le$ 100~MeV 
and are compatible with a most probable value $L\approx 60$~MeV, roughly 
corresponding to a power-law coefficient $\gamma = 0.6$.\cite{lipr08,tsang09,baoan11,lattlim12} 
The full line in Fig.~\ref{fig:params} represents, e.g., the result deduced by Li and Chen
from the MSU isospin-diffusion data and the neutron-skin thickness in $^{208}$Pb.\cite{lichen05}
The corresponding slope parameter is $L=65$~MeV. 
Rather similar constraints have been deduced from very recent investigations and observations 
of neutron-star properties.\cite{steiner12,hebe10,steiner10}

Considerable progress regarding the correlation of the symmetry energy
with particular observables in different models
has been made by the Florida and Barcelona 
groups.\cite{todd_piek05,rocamaza11} In continuation of the work of Typel and Brown
and of Furnstahl,\cite{typel01,furn02} a universal 
correlation between the thickness of the neutron-skin of $^{208}$Pb and the $L$ parameter 
has been found for empirical mean-field interactions.\cite{rocamaza11} 
The determination of the neutron-skin thickness by measuring the 
parity-violating contribution to electron scattering at high energy will thus offer a practically
model-free access to the slope at saturation, even though it is obtained by probing  
nuclear matter at an average density of typically 2/3 of this value.\cite{horo12}

\section{Dynamics of Heavy-Ion Reactions}

Densities of two to three times the saturation density may be reached on time scales 
of $\approx 10 - 20$~fm/c in the central zone of heavy-ion collisions at
relativistic energies of up to $\approx 1$~GeV/nucleon.\cite{li_npa02} 
The resulting pressure produces a collective outward motion of the compressed 
material whose strength will be influenced by the symmetry energy in asymmetric 
systems.\cite{dani02} At the same time, the excitation of $\Delta$ resonances in hard 
nucleon-nucleon scatterings leads to the production and subsequent emission of 
charged and neutral $\pi$ and $K$ mesons. 
The relative intensities of isovectors pairs of mesons depend directly on the 
proton-neutron content of the site where they are produced,\cite{stock86} 
suggesting them as sensitive probes for the high-density symmetry energy. 

Measurements of K$^+$/K$^0$ production ratios have been performed by the FOPI 
Collaboration for the neutron-rich and neutron-poor $A=96$ systems 
$^{96}$Zr + $^{96}$Zr and $^{96}$Ru + $^{96}$Ru.\cite{xlopez07} 
The experimental acceptances for these particles were considerably different, because of
their different decay channels, and the precision required for an isotopic study was more
reliably achieved with double ratios obtained from the data for the two systems with
different isotopic compositions. 
A significant sensitivity to the symmetry energy was expected from calculations under
equilibrium conditions but was found to be considerably reduced when 
actual collisions were modeled.\cite{xlopez07}  

The FOPI Collaboration has also presented an extensive account of the $\pi^-/\pi^+$ 
production ratios measured for the reaction $^{197}$Au + $^{197}$Au at several
energies up to 1.5 GeV/nucleon.\cite{reis07} 
Theoretical analyses of these data, however, come to
rather conflicting conclusions, suggesting everything from a rather stiff to a
super-soft behavior of the symmetry energy.\cite{ferini05,xiao09,feng10} 
The reason may lie in the treatment of the $\Delta$ dynamics and in competing effects 
of the mean fields and the $\Delta$ thresholds which will
require further studies.\cite{ferini05}
In the analysis of Xiao et {\it et al.} with the IBUU04 transport model,\cite{xiao09} 
the measured system and impact-parameter dependences have been reproduced,
however, only by assuming a super-soft dependence of the symmetry energy on density, close to
the $x=+1$ case shown in Fig.~\ref{fig:params}.
The presently inconclusive situation for pion ratios is very unfortunate because 
variations of up to 20\% for soft versus stiff parameterizations are expected. 

Dynamical flow observables have been proposed by several groups as probes 
for the equation of state at high density,\cite{li02,greco03} 
among them the so-called differential neutron-proton 
flow which is the difference of the parameters describing the collective motion of free 
neutrons and protons weighted by their numbers.\cite{li02} As pointed out 
by Yong {\it et al.},\cite{yong06} this observable minimizes the influence of the isoscalar 
part in the EoS while maximizing that of the symmetry term. Its proportionality to the particle
multiplicities, however, makes its determination very dependent on the experimental 
efficiencies of particle detection and identification and on the precise procedure for 
distinguishing free and bound nucleons in calculations. Therefore, also flow 
differences or ratios have been considered. 

\begin{figure}[!htb]
 \leavevmode
 \begin{center}
  \includegraphics[width=0.64\columnwidth]{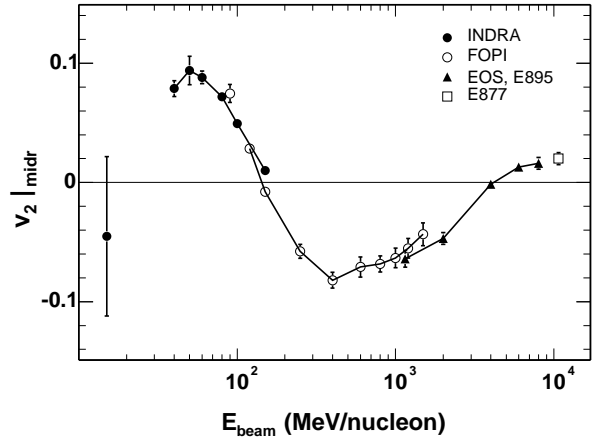}
 \end{center}
\vskip -0.1cm  
\caption{Elliptic flow parameter $v_{2}$ at mid-rapidity for $^{197}$Au+$^{197}$Au 
collisions at intermediate impact parameters (about 5.5-7.5 fm) as a function of incident
energy, in the beam frame. The filled and open circles represent the INDRA
and FOPI data,\protect\cite{lukasik05,andronic05} respectively, for $Z=1$ particles,
the triangles represent the EOS and E895 data\protect\cite{pinkenburg99} for
protons and the square represents the E877 data\protect\cite{bmunzinger98}
for all charged particles
(from Ref.\protect\cite{andro06},
reprinted with kind permission from Springer Science+Business Media).
}
\label{fig:v2corr}
\end{figure}

\section{Elliptic Flow}

The so-called squeeze-out of nuclear matter from the compressed interaction zone has 
first been observed in experiments at the Bevalac.\cite{gutbrod90} In a
sphericity analysis, the event shape in three dimensions was characterized by a
kinetic-energy flow tensor whose main orientation with respect to the beam direction
represents the collective sidewards flow and whose cross section, if it is non-isotropic,
indicates the existence of elliptic flow. At the bombarding energies of up to
1.05 GeV/nucleon investigated in these studies, a preferential emission of charged particles
perpendicular to the reaction plane has been observed. 
The shadowing by the spectator remnants as they pass
each other during the collision reduces the in-plane flow, so that the strength of the
off-plane emission, as quantified by the azimuthal anisotropy, reflects the internal pressure.

It has become customary to express both, directed and elliptic flows, 
and possibly also higher flow components by means of a Fourier decomposition
of the azimuthal distributions measured with respect to the orientation of the 
reaction plane $\phi_R$,\cite{voloshin96,ollitrault97,poskanzer98}   
\begin{equation} 
\frac{dN}{d(\phi-\phi_R)} = \frac{N_0}{2\pi} 
\left( 1+2 \sum_{n\geq1} v_n \cos n(\phi-\phi_R)\right),
\label{eq:defvn} 
\end{equation}

\noindent where $N_0$ is the azimuthally integrated yield. The
coefficients $v_{n} \equiv \langle\cos n(\phi-\phi_R)\rangle$ are functions of 
particle type, impact parameter, rapidity $y$, and the transverse momentum $p_t$; 
$v_{1}$ and $v_{2}$ represent the directed and elliptic flows, respectively.

Elliptic flow has become an important observable at other energy regimes as well. At
ultrarelativistic energies, the observation of the constituent-quark scaling of elliptic flow
is one of the prime arguments for deconfinement during the early collision phase,
and properties of the formed quark-gluon liquid are deduced from the observed
magnitude of collective motions.\cite{abelev07,fries08,snell11,adare12}
It implies that elliptic flow develops very early in the collision which is valid also
in the present range of relativistic energies, as confirmed by calculations.\cite{dani00}
Isotopic flow differences appear thus very suitable for studying
mean-field effects at high density.

An excitation function of the elliptic flow of $Z=1$ particles in 
$^{197}$Au+$^{197}$Au collisions from various experiments 
is shown in Fig.~\ref{fig:v2corr}. Squeeze-out perpendicular to the reaction plane, 
i.e. $v_2 < 0$, as a result of shadowing by the spectator remnants is observed
at incident energies between about 150~MeV/nucleon and 4~GeV/nucleon with a maximum near
400 MeV/nucleon.  At lower energies, the collective angular momentum in the mean-field 
dominated dynamics cause the observed in-plane enhancement of emitted reaction products.
The figure also shows that elliptic flow can be measured quite precisely. The
reliability of the applied methods is being demonstrated by the good agreement of data sets from
different experiments in the overlap regions of the studied intervals in collision 
energy.\cite{lukasik05,andro06,reisdorf12} 

\begin{figure}[htb!]
\centering
\includegraphics*[width=64mm]{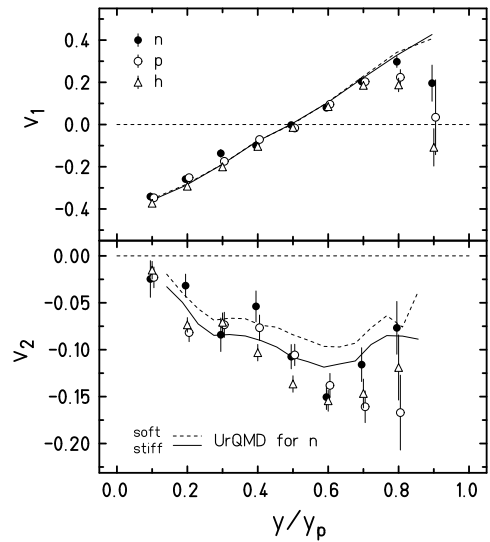}
\vskip -0.1cm
\caption{Measured flow parameters $v_1$ (top) and $v_2$ (bottom)
for mid-peripheral (5.5~$\le b \le$~7.5 fm) $^{197}$Au + $^{197}$Au collisions at 400 MeV/nucleon
for neutrons (dots), protons (circles), and hydrogen isotopes ($Z=1$, open triangles) 
integrated within $0.3 \le p_t/A \le 1.3$~GeV/c/nucleon
as a function of the normalized rapidity $y_{\rm lab}/y_p$. 
The UrQMD predictions for neutrons are shown for the FP1 parameterization of the in-medium 
cross sections and for a stiff ($\gamma = 1.5$, full lines) and a 
soft ($\gamma = 0.5$, dashed) density dependence of the symmetry term.  
The experimental data have been 
corrected for the dispersion of the reaction plane 
(reprinted from Ref.\protect\cite{russotto11}, Copyright (2011), with permission from Elsevier).
}
\label{fig:fig3}
\end{figure}

\section{Results from FOPI/LAND}

The squeeze-out of neutrons has first been observed by the FOPI/LAND Collaboration who
studied the reaction $^{197}$Au + $^{197}$Au at 400 MeV/nucleon.\cite{leif93}
The squeeze-out of charged particles reaches its maximum at this energy (Fig.~\ref{fig:v2corr}),
and similarly large anisotropies were observed for neutrons.\cite{lamb94}
The neutrons had been detected with the Large-Area Neutron Detector, LAND,\cite{LAND} 
while the FOPI Forward
Wall, covering the forward hemisphere of laboratory angles $\theta_{\rm lab} \le 30^{\circ}$ 
with more than 700 plastic scintillator elements, was used to determine the
modulus and azimuthal orientation of the impact parameter.

The motivation for returning to the existing data set has been provided by studies 
performed with the UrQMD transport code for this fairly neutron-rich system ($N/Z = 1.49$) 
which indicated a significant sensitivity of the elliptic-flow parameters to the 
assumptions made for the density dependence of the symmetry energy.\cite{russotto11} 
In calculations with power-law coefficients $\gamma = 0.5$ and 1.5 
(cf. Eq.~\ref{eq:pot_term} and Fig.~\ref{fig:params}), 
the relative strengths of neutron
and proton elliptic flows were found to vary on the level of 15\%.

The performed reanalysis of the data consisted mainly in choosing 
equal acceptances for neutrons and hydrogen isotopes with regard to 
particle energy, rapidity and transverse momentum (energy and momentum 
per nucleon for deuterons and tritons). The results obtained for a 
mid-peripheral event class are shown in Fig.~\ref{fig:fig3} as a
function of the rapidity $y$, normalized with respect to the projectile rapidity 
$y_p = 0.896$. 
Their asymmetry with respect to mid-rapidity, $y/y_p=0.5$, is caused
by the LAND detector whose kinematic acceptance in $p_t$ increases with $y$. 
With the yields dropping at large $p_t$, the statistical errors become large at forward rapidity.
The theoretical predictions have been obtained simulating the LAND acceptance and the 
experimental analysis conditions. The results, shown for neutrons in Fig.~\ref{fig:fig3},
follow qualitatively the experimental data.

\begin{figure}[htb!]
\centering
   \includegraphics*[width=60mm]{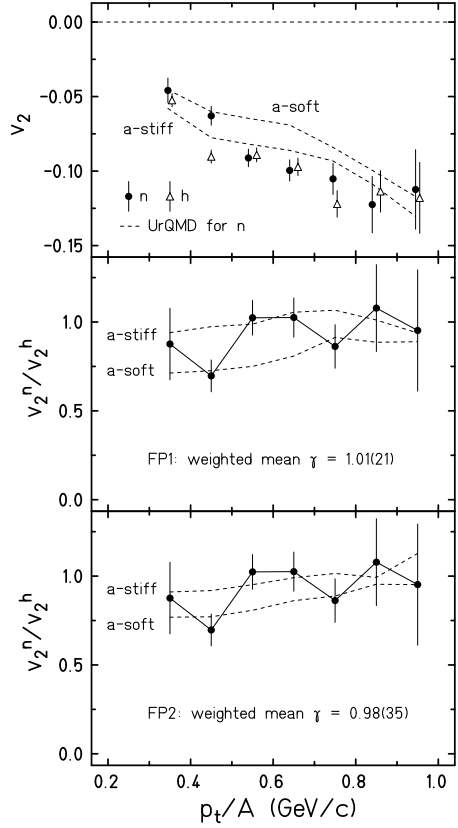}
\vskip -0.1cm
\caption{Elliptic flow parameters $v_2$ for neutrons (dots) and 
hydrogen isotopes (open triangles, top panel) and their ratio (lower panels) for moderately 
central ($b<7.5$ fm) collisions of $^{197}$Au + $^{197}$Au at 400 MeV/nucleon, integrated
within the rapidity interval $0.25 \le y/y_p \le 0.75$, as a function 
of the transverse momentum per nucleon $p_t/A$. The symbols represent the experimental data.
The UrQMD predictions for $\gamma = 1.5$ (a-stiff) and $\gamma = 0.5$ (a-soft)
obtained with the FP1 parameterization for neutrons (top panel) and for 
the ratio (middle panel),
and with the FP2 parameterization for the ratio (bottom panel)
are given by the dashed lines 
(reprinted from Ref.\protect\cite{russotto11}, Copyright (2011), with permission from Elsevier).
}
\label{fig:fig4}
\end{figure}

According to the UrQMD model, the sensitivity of the directed flow of neutrons to the stiffness 
of the symmetry energy is nearly negligible (Fig.~\ref{fig:fig3}, top panel) while it is
significant for the elliptic flow (bottom panel). 
This is evident also from the dependence of the elliptic flow parameter $v_2$ on the transverse 
momentum per nucleon, $p_t/A$, shown in Fig.~\ref{fig:fig4}, upper panel, for the full 
statistics of central and mid-peripheral collisions ($b \le 7.5$~fm) collected in this
experiment. The measured values 
are approximately reproduced by the UrQMD predictions which are significantly 
different for the stiff ($\gamma=1.5$) and soft ($\gamma=0.5$) density dependences. 

\section{Elliptic-Flow Ratios}

For the quantitative evaluation, the ratio of the flow parameters of neutrons versus protons 
or versus $Z=1$ particles has been proposed as a useful observable.\cite{russotto11}  
Systematic effects influencing the collective flows of neutrons and charged particles 
in similar ways should thereby be minimized, on the experimental as well as on the 
theoretical side. These include, e.g., the existing uncertainties of the isoscalar EoS 
which will affect the absolute magnitude of calculated flows and 
the matching of the impact-parameter intervals used in the 
calculations with the corresponding experimental event groups. To demonstrate this 
kind of insensitivity,
the calculations were performed with two parameterizations of the momentum dependence
of the elastic nucleon-nucleon cross section, labelled FP1 and FP2, which differ in their
absolute predictions of $v_2$ at mid-rapidity by $\approx 40\%$ for this 
reaction.\cite{qli10,qli11}

In Fig.~\ref{fig:fig4} (lower panels), the results for the ratio with respect to the 
total hydrogen yield are shown. 
The calculated ratios exhibit clearly the sensitivity of the elliptic flow to the stiffness 
of the symmetry energy predicted by the UrQMD but depend only weakly on the chosen
parameterization for the in-medium nucleon-nucleon cross section. 
The experimental ratios, even though associated with large errors, 
scatter within the interval given by the two calculations.
Linear interpolations between the predictions, averaged over  
$0.3 < p_t/A \le 1.0$ GeV/c,
yield very similar results $\gamma = 1.01 \pm 0.21$ and $\gamma = 0.98 \pm 0.35$ for 
the two parameterizations. The error is larger for FP2 because the sensitivity is somewhat
smaller (Fig.~\ref{fig:fig4}, bottom panel).

\begin{figure}[!htb]
 \leavevmode
 \begin{center}
  \includegraphics[width=0.55\columnwidth]{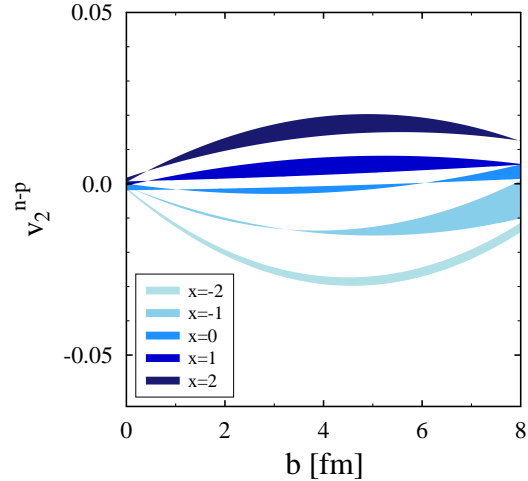}
 \end{center}
\vskip -0.1cm  
  \caption{Sensitivity of the neutron-proton elliptic flow difference 
$v_2^{n-p} = v_2^n - v_2^p$ ($v_2$ as defined in Eq.~\protect\ref{eq:defvn}) 
to the isospin independent EoS and to 
the symmetry term in the parameterization of Ref.\protect\cite{lichen05} as
indicated. The widths of the bands represent the uncertainty arising from using either a 
soft ($K = 210$~MeV with momentum dependence) or a hard ($K = 380$~MeV) version for the
isospin-independent part of the EoS 
(reprinted from Ref.\protect\cite{cozma11}, Copyright (2011), with permission from Elsevier).
}
\label{fig:cozma}
\end{figure}

This analysis was repeated in various forms. With the squeeze-out ratios $v_2^n/v_2^p$ of 
neutrons with respect to free protons, similar results were obtained, however with larger errors.
The study of the impact parameter dependence indicated a slightly smaller value 
$\gamma \approx 0.5$ for the mid-peripheral event group, again with larger errors. 
It was also tested to which density region around $\rho_0$ the elliptic-flow
ratios are sensitive, with the result that both,
sub- and supra-saturation densities are probed with this observable.\cite{russotto11}

In consideration of the apparent systematic and experimental errors, a value
$\gamma = 0.9 \pm 0.4$ has been adopted by the authors as best representing the power-law 
exponent of the potential term resulting from the elliptic-flow analysis. 
It falls slightly below the $\gamma = 1.0$ line shown in Fig.~\ref{fig:params} but, with the
quoted uncertainty, stretches over the interval from $\gamma = 0.5$ halfway up to $\gamma = 1.5$. 
The corresponding slope parameter is $L = 83 \pm 26$~MeV.
The squeeze-out data thus indicate a moderately soft to linear
behavior of the symmetry energy that is consistent with the density dependence 
deduced from experiments probing nuclear matter near or below saturation.
Comparing with the many-body
theories shown in Fig.~\ref{fig:fuchs06b}, the elliptic-flow result is in good qualitative 
agreement with the range spanned by the DBHF and variational calculations based on realistic
nuclear potentials.

\section{Elliptic-Flow Differences}

In an independent analysis, Cozma has used data from the same experiment and investigated 
the influence of several parameters on the difference  
between the elliptic flows of protons and neutrons 
using the T\"{u}bingen version of the QMD transport model.\cite{cozma11} They included the
parameterization of the isoscalar EoS, the choice of various forms of free or in-medium
nucleon-nucleon cross sections, and model parameters as, e.g., the widths of the wave 
packets representing nucleons. 
The interaction developed by Das {\it et al.} was used which contains an explicit 
momentum dependence of the symmetry energy part.\cite{das03,lichen05} 
In Fig.~\ref{fig:cozma}, the difference of neutron and proton elliptic flows for five
choices of the $x$ parameter describing the density dependence of the symmetry energy
is reproduced from this work. It shows the large sensitivity to the stiffness of the 
symmetry energy (different $x$-parameters) and the reduced influence of the isoscalar EoS 
(widths of the bands).
A super-soft behavior of the symmetry energy was confirmed to be excluded by the comparison 
with the experimental flow data. 

Because the flow difference is still proportional to the overall magnitude of the
elliptic flow predicted by the model, also the ratios of the neutron-versus-proton
elliptic-flow parameters have been calculated and compared to
the FOPI/LAND data.\cite{cozma_private} 
The best description is obtained with a density dependence slightly stiffer than the
$x=0$ solution which approximately corresponds to the $\gamma=0.5$ case shown
in Fig.~\ref{fig:params}.
It is thus rather close to the UrQMD result 
$\gamma = 0.9$ 
which may represent an important step towards the model invariance that one would 
ultimately like to achieve, 
not only because of the invariance with respect to the treatment of the 
nucleon-nucleon cross sections observed in the two studies 
but also because of the explicit momentum dependence of the isovector potentials which
is implemented in the T\"{u}bingen QMD but not in the UrQMD version used.

\section{Conclusion and Outlook}

According to the predictions of transport models, the relative strengths of
neutron and proton elliptic flows represent an observable sensitive to the symmetry
energy at densities near and above saturation. By forming ratios or differences
of neutron versus proton or neutron versus hydrogen flows, the influence of isoscalar-type
parameters of the model descriptions can be minimized.

The comparison of the results obtained from the FOPI/LAND data for $^{197}$Au + $^{197}$Au 
reactions at 400 MeV/nucleon with UrQMD transport calculations favors a moderately soft to 
linear symmetry term with a density dependence
of the potential term proportional to $(\rho/\rho_0)^{\gamma}$ with $\gamma = 0.9 \pm 0.4$,
compatible with predictions of ab-initio calculations.
With the T\"{u}bingen QMD model, a similar result is obtained.

The explicit proof that the elliptic-flow ratios are probing the nuclear mean field at 
super-saturation densities is, so far, limited to test calculations made with the 
UrQMD.\cite{russotto11} They indicate that the strength of the symmetry energy at 
densities both, below and above saturation, are essential. 
It will be useful to study this in more detail.

The statistical uncertainty of the existing FOPI/LAND data is larger than the investigated
systematic effects of model parameters and analysis techniques. More significant data 
can thus be expected from a new experiment delivering a comprehensive data set
with sufficiently high statistical accuracy. Explicitly testing the predicted dependences on
rapidity and transverse momentum will provide useful constraints for the models. Additional
observables as, e.g., neutron-to-proton or $^3$H-to-$^3$He yield ratios may serve in
resolving ambiguities caused by the effective-mass splitting between neutrons and protons. 
It will be interesting to see first results appearing from the recently completed measurements 
of the ASYEOS Collaboration at the GSI Laboratory.\cite{s394}

Because of the quadratically rising importance of the symmetry energy, the continuation
of this program with systems of larger asymmetry is very promising and
important. The lower luminosities to be expected from the use of secondary beams
and isotopically enriched targets will have to be compensated with efficient 
detector setups. 
The neutron detector NeuLAND proposed for experiments at FAIR will offer a highly improved
detection efficiency for neutrons in the energy range 100 to 400 MeV.\cite{NeuLAND} 
This will be essential for 
extending the program also to reactions at lower energies for which significant mean-field 
effects are predicted for directed and elliptic flows.\cite{guo12}

Stimulating and fruitful discussions with M.D.~Cozma, M.~Di~Toro, 
W.~Reisdorf, and with the authors of Ref.\cite{russotto11}, P.~Russotto, P.Z.~Wu, M.~Zoric, 
M.~Chartier, Y.~Leifels, R.C.~Lemmon, Q.~Li, J.~{\L}ukasik,
A.~Pagano, and P.~Paw{\l}owski are gratefully acknowledged.


\end{document}